# Gated two-dimensional electron gas in magnetic field: nonlinear versus linear regimes

N. Dyakonova [1], M. Dyakonov [1, 2], and Z.D. Kvon [3, 4]

[1] Charles Coulomb Laboratory, University of Montpellier, 34095 Montpellier, France
[2] Ioffe Institute, 194021 St. Petersburg, Russia
[3] Institute of Semiconductor Physics, 630090 Novosibirsk, Russia
[4] Novosibirsk State University, 630090 Novosibirsk, Russia

**Abstract.** We study the effect of magnetic field on the properties of a high mobility gated two-dimensional electron gas in a field effect transistor with the Hall bar geometry. When approaching the current saturation when the drain side of the channel becomes strongly depleted, we see a number of unusual effects related to the magnetic field induced re-distribution of the electron density in the conducting channel. The experimental results obtained in the non-linear regime have been interpreted based on the results obtained in the linear regime by a simple theoretical model, which describes quite well our observations.

## I. INTRODUCTION

We study the effect of magnetic field on the properties of a high mobility gated two-dimensional electron gas in a Field Effect Transistor (FET) with the Hall bar geometry. In the linear regime, we observe the well-known magnetoresistance and Hall effect which were previously thoroughly studied in many works, see e.g.[1-11]. However, when approaching the current saturation in which the drain side of the channel becomes strongly depleted, we see a number of unusual effects related to the magnetic field induced re-distribution of the electron density in the conducting channel.

The standard expressions for the components of electric current density $\boldsymbol{j}$ in magnetic field are:

$$j_x = \sigma_{xx}E_x + \sigma_{xy}E_y, \qquad (1)$$
$$j_y = \sigma_{yx}E_x + \sigma_{xx}E_y, \qquad (2)$$

where $\sigma_{ik} = en\mu_{ik}$ is the conductivity tensor, $\sigma_{xx} = \sigma_{yy}$, $\sigma_{yx} = -\sigma_{xy}$ , $e$ is the electron charge, $n$ is the 2D electron concentration, $\mu_{ik}$ is the mobility tensor, $E_x = -\partial V/\partial x$ and $E_y = -\partial V/\partial y$ are the components of the electric field, $V(x, y)$ is the electrostatic potential in the channel, which (due to the condition div $j$ = 0) satisfies the Laplace equation.

For a Hall bar with length $L$ and width $w$ the boundary conditions correspond to fixed potentials $V_s$ and $V_d$ at the source ($x$ = 0) and the drain ($x$ = $L$), respectively, and the absence of current $j_y$ through the lateral boundaries at $y = \pm w/2$.

In the presence of magnetic field $B$, the mobility tensor $\mu_{ik}$ generally depends on $B$, the simplest situation being described by the well-known Drude-like expressions:

$$\mu_{xx} = \mu_{yy} = \mu \frac{1}{1+\beta^2}, \quad \mu_{xy} = -\mu_{yx} = \mu \frac{\beta}{1+\beta^2}, \qquad (3)$$

where $\mu$ is the mobility in the absence of magnetic field, and $\beta = \Omega\tau$, $\Omega = eB/mc$ is the cyclotron frequency, $\tau$ is the relaxation time, $m$, $c$ are the electron effective mass and light velocity, respectively.

It is well known for a long time that in this simplest case the effect of magnetic field is reduced to the Hall effect and to the distortion of the equipotential lines in the vicinity of the source and drain contacts (Fig. 1), which results in the so-called *geometrical* magnetoresistance [2–4], which depends on the ratio $w/L$ and the dimensionless magnetic field $\beta$ (or, equivalently, the Hall angle).

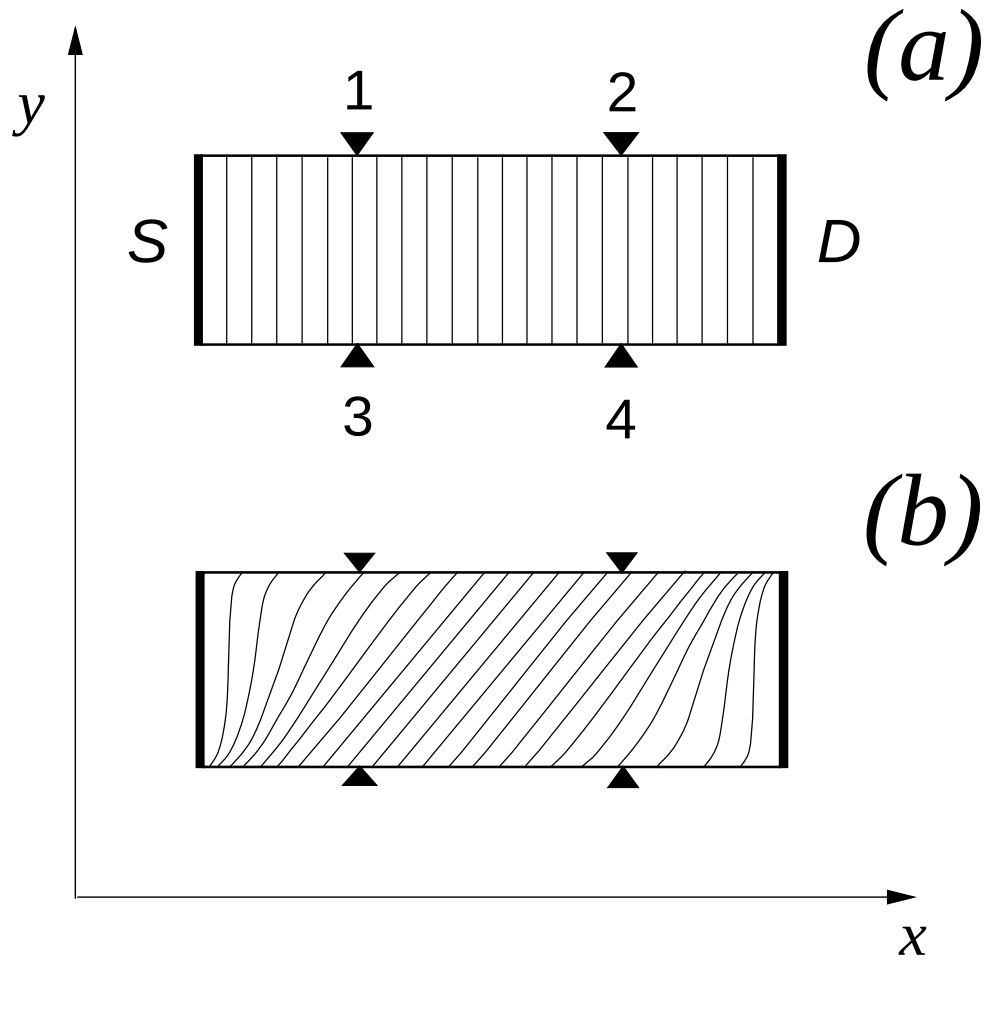

FIG. 1: Schematics of the equipotential lines for a Hall bar with source ($S$) and drain ($D$) contacts: ($a$)- at zero magnetic field, ($b$) - in the presence of magnetic field. Side contacts for measuring Hall and longitudinal voltages are shown. In our samples, the contacts 1, 3 and 2, 4 are situated symmetrically with respect to the middle of the sample.

There are also many reasons for the existence of a *physical* magnetoresistance [5–11] due to the magnetic field dependence of the mobility tensor $\mu_{ik}$ different from that given by the simple formulas in Eq. (3).

Here, our main goal is *not* in defining, nor studying the mechanisms of magnetoresistance, but rather, in *relating* the effects of magnetic field in the linear and nonlinear regimes, in other words, in understanding the observed influence of magnetic field in the nonlinear regime on the basis of our experimental results obtained in the linear regime.

## II. THEORY: LINKING THE EFFECTS OF MAGNETIC FIELD IN THE LINEAR AND NONLINEAR REGIMES

The specific feature of the gated electron gas in a FET (see Fig. 2), is that the local electron concentration, *n*, is controlled by the so-called gate voltage swing *V* by the plane capacitor law [12]:

$$en = CV, \tag{4}$$

where *en* is the charge density in the channel, *C* is the gate-to-channel capacitance per unit area, $V(x, y) = V_{ch}(x, y) - V_{th}$ is the local gate voltage swing, $V_{ch}(x, y)$ is the local channel potential with respect to the gate, $V_{th}$ is the threshold voltage. For $V \leq 0$ the channel does not contain any electrons.

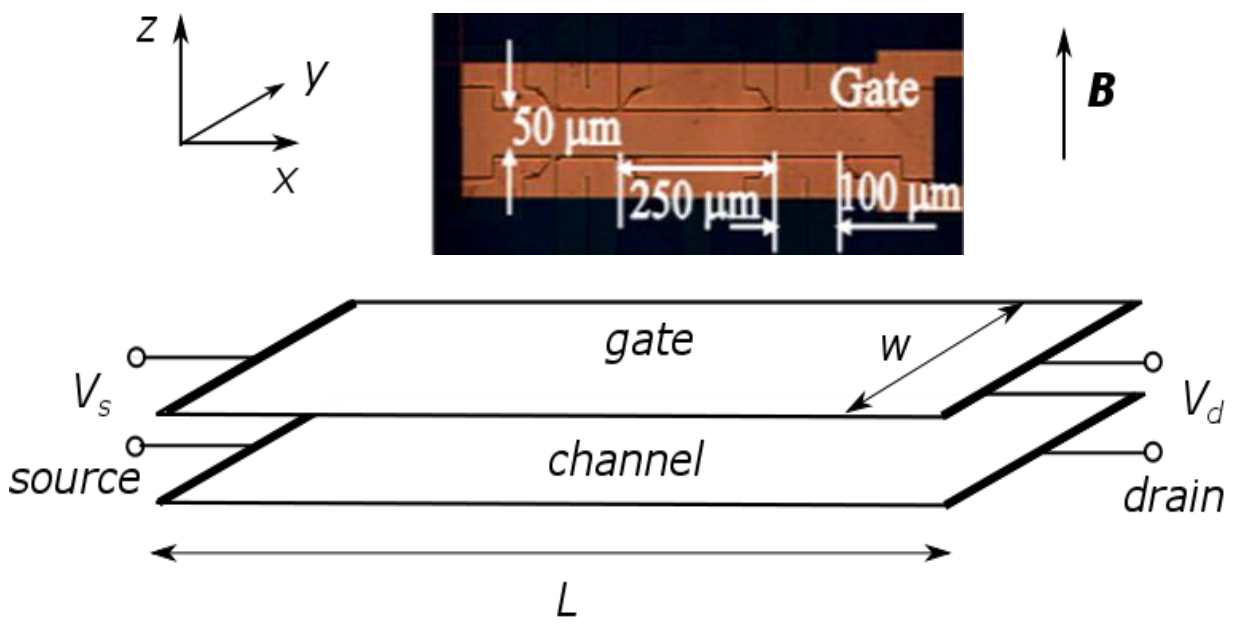

FIG. 2: The image of our sample and schematics of a FET with the Hall bar geometry, $w/L \approx 0.1$.

As a consequence of Eq. (4), the basic Eqs. (1), (2), as applied to a gated electron gas, become nonlinear, since the conductivity tensor $\sigma_{ik} = en\mu_{ik}$ is now proportional to the local voltage swing *V*. Thus

$$\sigma_{xx}\frac{\partial V}{\partial x} = \mu_{xx}CV\frac{\partial V}{\partial x} = en_s\mu_{xx}\frac{\partial}{\partial x}\left(\frac{V^2}{2V_s}\right), \tag{5}$$

and similar for $\sigma_{xy}\partial V/\partial y$. Here $V_s$ is the gate voltage swing at the source, $n_s = CV_s/e$ is the electron concentration at the source. Note that the current density in a gated electron gas is proportional to the gradient of the *square* of the gate voltage swing *V*.

It is remarkable, that Fig. 1 remains valid for the gated electron gas, *but* with the important difference that now the lines correspond to fixed values of the *square* of potential $V^2$, rather than of the potential *V* itself, like in the linear regime where the electron concentration is fixed. The difference becomes important when the variation of *V* in the channel becomes strong enough, i.e. in the nonlinear regime, especially when approaching the current saturation caused by the depletion of the drain side of the channel.

In the absence of magnetic field, this non-linearity is well known [12] and is described by a simple equation for the drain current $I = jw$ following from Eqs. (1-5) for $B = 0$, $E_y = 0$:

$$I = \frac{V_{sd}}{R}\left(1 - \frac{V_{sd}}{2V_s}\right) = \frac{1}{R}\frac{V_s^2 - V_d^2}{2V_s}, \tag{6}$$

where $V_{sd} = V_s - V_d$ is the voltage applied between the source and the drain, $V_s$ and $V_d$ are the gate voltage swings at the source and the drain respectively, *R* is the resistance of the two-dimensional slab with electron concentration $n_s = CV_s/e$. This equation is valid up to $V_{sd} = V_s$ (or $V_d = 0$), when the channel becomes strongly depleted at the drain and the current saturates. With further increase of $V_{sd}$ the current remains constant while the mechanism of conduction in the drain region changes as described in Ref. 12.

We now come to our main point: establishing a simple relationship between the effects of magnetic field in the linear ($V_{sd} << V_s$) and non-linear ($V_{sd} \sim V_s$) regimes.

As shown above, see Eqs. (5,6), for a gated electron gas in magnetic field, the potential *V* in Eqs. (1, 2), as well as in the boundary conditions, should be replaced by $V^2/(2V_s)$, similar to Eq.(5), and this is our main idea. After this replacement, it will follow from Eqs. (1, 2) that now the quantity $V^2/(2V_s)$ must satisfy the Laplace equation (equivalent to div $\boldsymbol{j} = 0$) with the same boundary conditions as for the potential *V* in an ungated electron gas.

This means that the result obtained for the magnetoresistance *R*(*B*) in the linear regime

$$I = \frac{V_s - V_d}{R(B)}, \tag{7}$$

can be directly applied to the non-linear regime by a simple replacement of *V* by $V^2/(2V_s)$.

Thus, for a gated electron gas we obtain:

$$I = \frac{V_s^2 - V_d^2}{2V_s R(B)} \tag{8}$$

which is equivalent to Eq. (6) with the constant resistance *R* replaced by *R*(*B*) *measured in the linear regime*.

We stress that this simple recipe is valid if the current density $\boldsymbol{j}$ is proportional to the *first power* of the electron concentration *n*, i.e. it does not apply to the case when Shubnikov-De Haas oscillations or the quantum Hall effect become important[1]. In other words, Eq. (8) strictly stands when the resistance *R*(*B*) is due to the combination of geometrical magnetoresistance and the dependence of the mobility tensor on magnetic field (but not on the electron

---

[1)] Those effects essentially depend on the position of the Fermi level, i.e. on the electron concentration

concentration). Our experimental results presented below belong to the regime where this condition is fulfilled (the amplitude of Shubnikov-De Haas oscillations is small).

The simple rule $V_s - V_a \rightarrow (V_s^2 - V_a^2)/(2V_s)$ for the transition from the linear to nonlinear regime in a gated electron gas should be equally valid for any side contact $a$. To practically apply this rule, one needs to determine in the linear regime the *"resistance"* $R_{sa} = (V_s - V_a)/I$, where $I$ is again the *drain* current. We stress, that $R_{sa}$ is *not* the resistance that could be determined by applying the voltage difference $V_{sa} = V_s - V_a$ and measuring the current in the *s-a* circuit. Rather, it is a coefficient that determines the voltage difference $V_{sa}$ induced by the *drain* current $I$.

Thus, in the nonlinear regime we have

$$I = (V_s^2 - V_a^2)/(2V_s R_{sa}), \qquad (9)$$

where the "resistance" $R_{sa}$ and its dependence on magnetic field can be established in the linear regime. From Eq. (9) we deduce the expression for $V_{sa}$ at arbitrary values of the drain current $I$:

$$V_{sa} = V_s\left(1 - \sqrt{1 - 2R_{sa}I/V_s}\right). \qquad (10)$$

For low drain currents, Eq. (10) reduces to $V_{sa} = IR_{sa}$.

## III. EXPERIMENT

The gated structures were fabricated on the base of the AlGaAs/GaAs heterostructure with two-dimensional electron gas (electron density $n_s = 6.5\cdot10^{11}$ cm$^{-2}$, mobility $\mu = 10^6$ cm$^2$/V·s ) by Ti/Au metallic layer evaporation (Ti layer of 20 nm thickness and Au layer of 100 nm thickness) on top of the AlGaAs/GaAs structure. The studied Hall bar shaped structures (length $L$=600 µm, width $w$=50 µm) were covered by a metallic gate 650 µm long and 150 µm wide. Side voltages were measured between contacts spaced by 250 µm and located at 200 µm from source and drain boundaries of the gated structure (see Fig. 2).

Ungated access zones of 2DEG between the gated structure and source and drain contact pads were 750 µm long and 100–200 µm wide. At $B$=0 the access zones resistance and the contact resistance at low drain voltage were estimated to be ~100 Ohm and ~ 950 Ohm respectively.

The presence of additional access and contact resistances practically does not influence the gate voltage swing $V_s$, neither in the linear regime, nor in the nonlinear regime.

Current-voltage characteristics and voltages were measured at 4.2 K in magnetic fields up to 2 Tesla using the Keithley source-meter and Agilent voltmeters.

In Fig. 3 we present the measured $I - V$ dependencies at $B$ = 0 for several values of the gate voltage $V_g$. The results are quite typical for gated structures: the current linearly increases with $V_{sd}$ and saturates when $V_{sd}$ approaches $V_s$, so that the condition $V_d$ = 0 at the drain side is achieved.

We remind that the usual notation $V_g$ for the gate voltage with respect to the source, habitually used by experimentalists, is linked to the value of $V_s$ as: $V_g = V_{th} + V_s$. The measured value of the threshold voltage is $V_{th} \approx -1.8$ V.

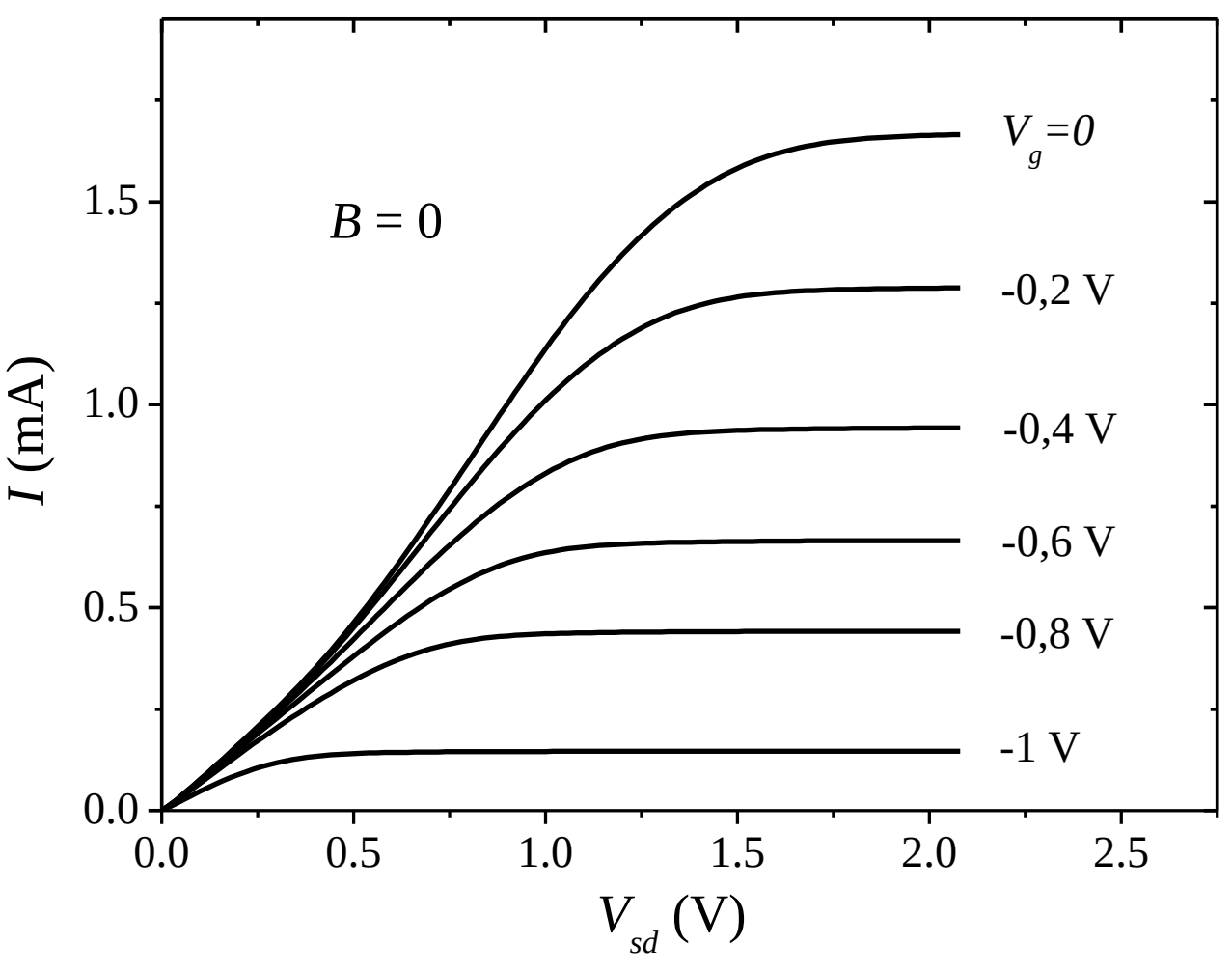

FIG. 3. Measured $I$ - $V$ dependencies at $B$ = 0 for different values of the gate voltage $V_g = V_{th} + V_s$.

### *A. Effects of magnetic field in the linear regime.*

We now present our experimental results for the magnetoresistance, the Hall effect, and the side voltages in the linear regime where $V_{sd} << V_s$, so that the electron density in the channel, $n \sim V_s$, is homogeneous.

#### *1. Magnetoristance*

Fig. 4($a$) presents the magnetoresistance $R(B)$ in the linear regime for several values of $V_g$. In Fig. 4($b$) we plot the magnetic field dependence of the product $R(B)\cdot V_s$ to check our assumption above that $R(B) \sim 1/n \sim 1/V_s$, i.e. that the mobility tensor does not depend on the electron concentration. It can be seen that for $B$ < 2T, to a reasonable approximation, this is indeed the case.

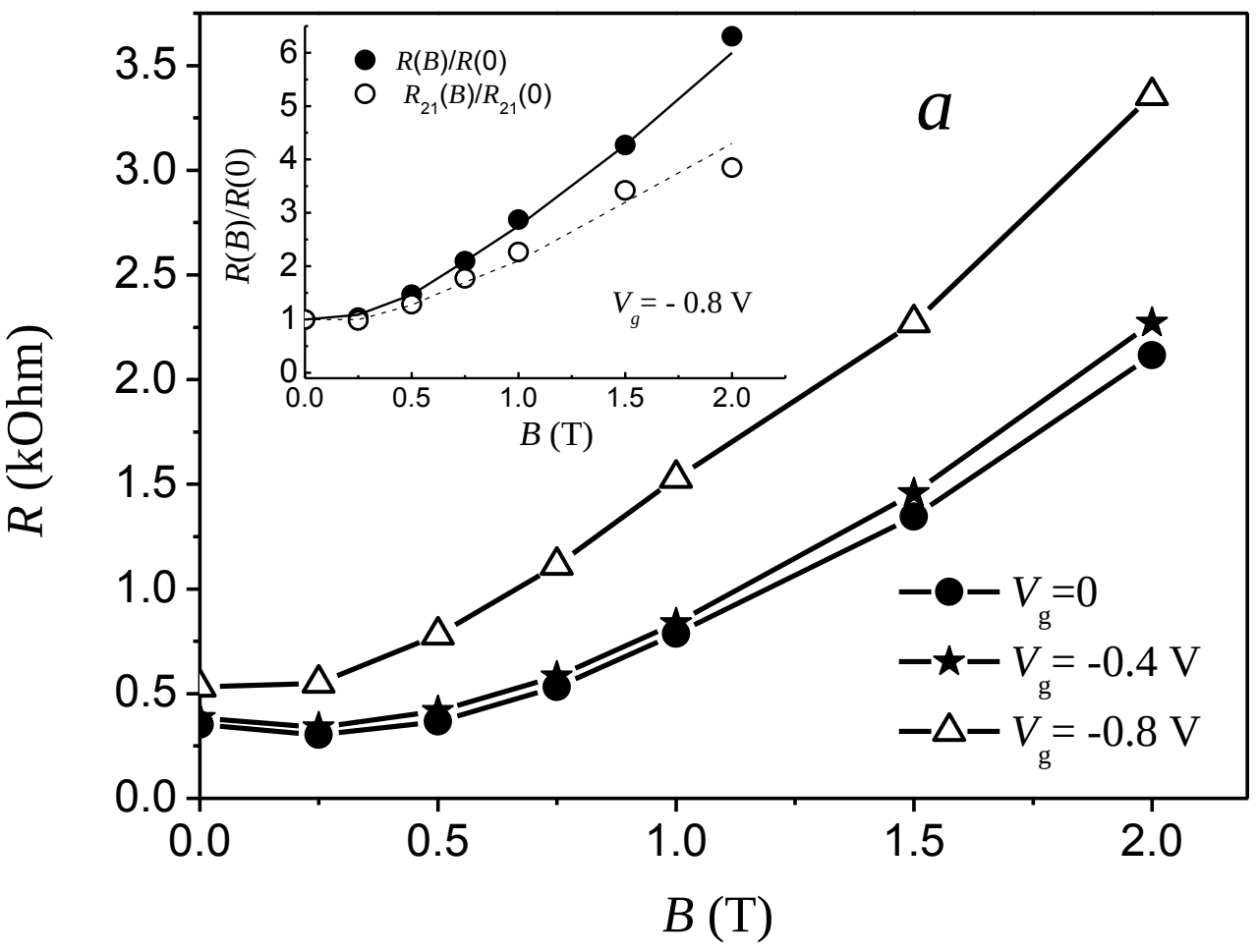

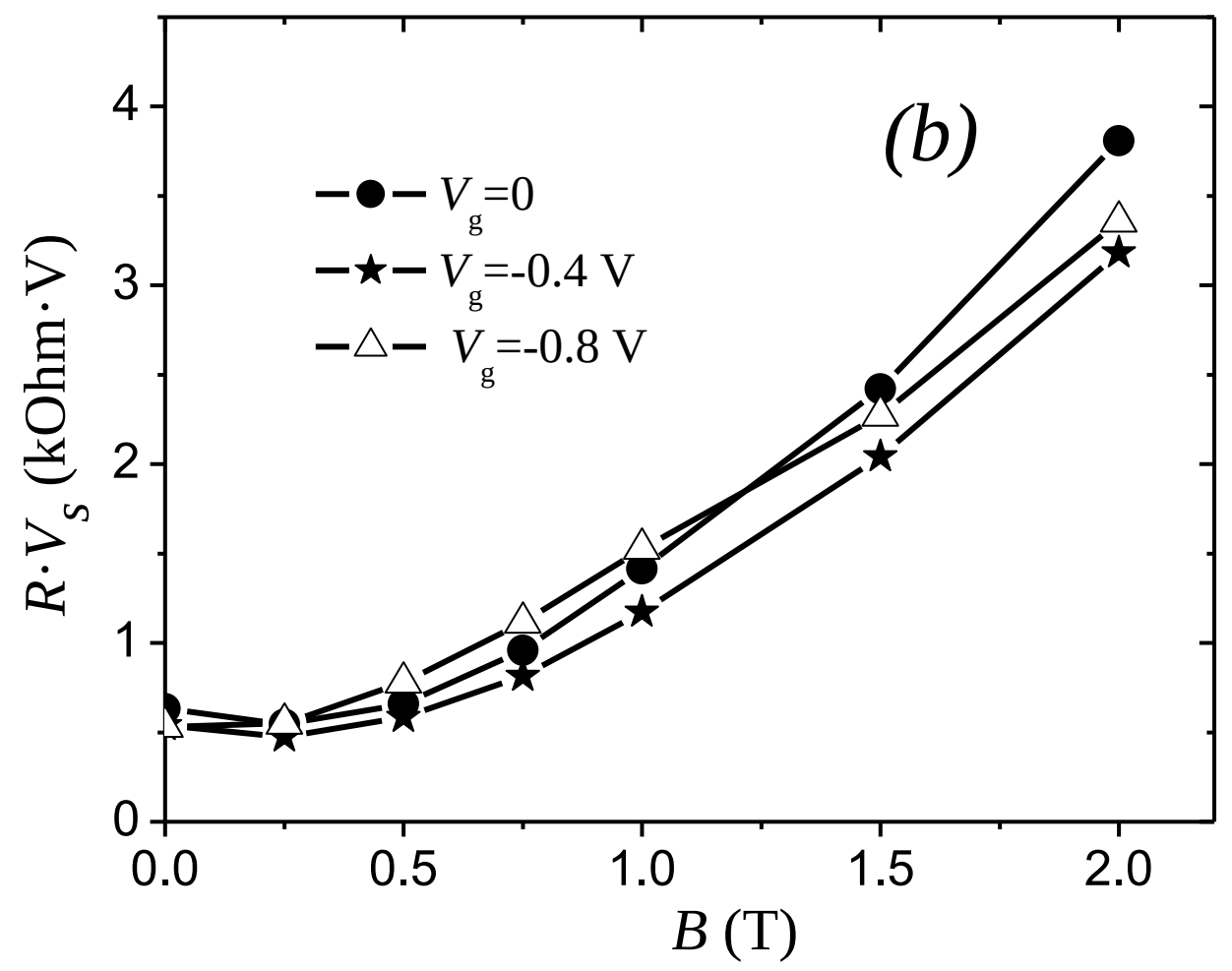

Fig. 4. (*a*) Magnetoresistance $R(B)$ in the linear regime ($V_{sd}$ = 0.1 V) for several values of $V_g$. Inset: $R(B)/R(0)$ and $R_{21}(B)/R_{21}(0)$ at $V_g$ =-0.8 V, $V_{sd}$ = 0.1V. (*b*) Product $R \cdot V_s$ as a function of $B$ for the same values of $V_g$.

To estimate the role of geometrical magnetoresistance which originates in the vicinity of the source and drain contacts [2, 3], we have measured the magnetic field dependence of the resistance $R_{21}$ between side contacts 2 and 1 (see inset to Fig. 4(*a*)), thus excluding the effect of geometrical magnetoresistance. Above 0.5 T the ratio $R_{21}(B)/R_{21}(0)$ increases nearly linear in magnetic field, similar to the total resistance.

This shows the existence of a *physical* magnetoresistance, which is linear in magnetic field. Positive magnetoresistance was previously observed experimentally [5–7] and discussed theoretically.[8–11]

Thus above ~ 0.5 T, the measured resistance $R(B)$ (i.e. the resistivity $\rho_{xx}$) increases linearly with $B$. On the other hand, the Hall resistivity $\rho_{xy}$ is also linear in $B$. This means that the topography of the equipotential lines becomes frozen above ~ 0.5 T.

### *2. Hall effect and side voltages*

In Fig. 5 we show the measured voltages at contacts 1,2,3,4 with respect to the source as functions of magnetic field. All voltages are linear in magnetic field up to ~ 1 T and saturate at higher fields. These measurements allow to extract the Hall voltages $V_{31}$ and $V_{42}$, as well as the side voltages $V_{12}$ and $V_{34}$. One can see that, as expected, $V_{31} = V_{42}$ and $V_{12} = V_{34}$ (this is no longer true in the nonlinear regime, see below).

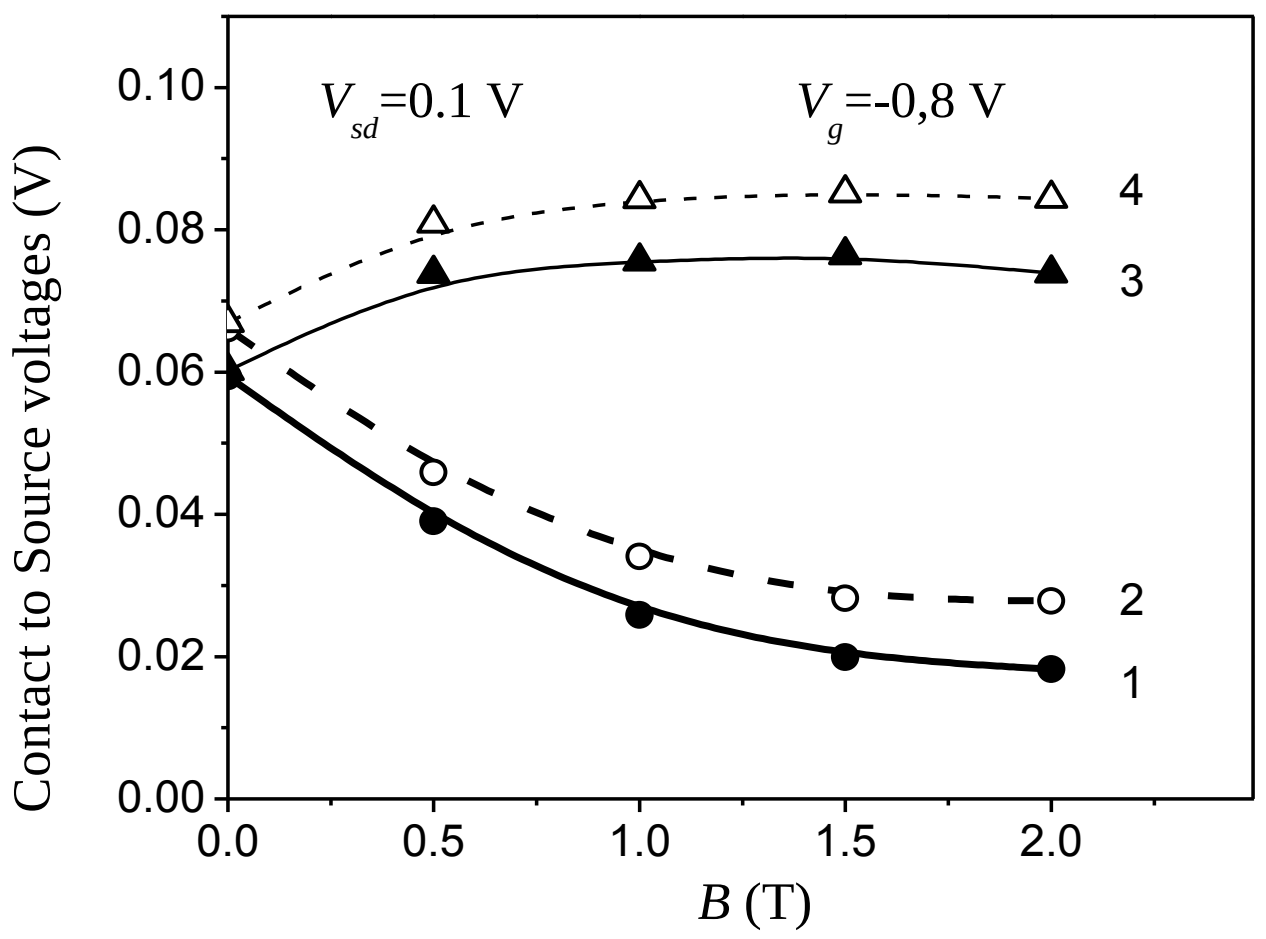

Fig. 5. Voltages measured between side contacts 1–4 and the source as functions of magnetic field $B$. Note the saturation at $B$ >1 T.

The experimentally observed saturation of all measured voltages at $B$ ~ 1T in the linear regime means that in our sample the pattern of equipotential lines defined by the ratio $\sigma_{xy}/\sigma_{xx}$ is frozen for higher magnetic fields. Whatever are the physical reasons for this, we must take into account this experimental fact when discussing the following results obtained in the nonlinear regime.

### ***B. Effects of magnetic field in the nonlinear regime***

#### *1. Magnetoristance*

The measured $I$–$V$ characteristics for different magnetic fields in the range 0 – 2 T are presented in Fig. 6(*a*). In Fig. 6(*b*), for the same values of magnetic field, we plot the ratio $I/I_{sat}$, where $I_{sat}$ is the saturation current for a given magnetic field. One can see that the dependence $R(B)$ in the nonlinear regime is indeed practically the same as in the linear regime, resulting in the remarkable merging of all the $I/I_{sat}$ *vs* $V$ curves for different values of $B$, as it is predicted by Eqs. (7) and (8).

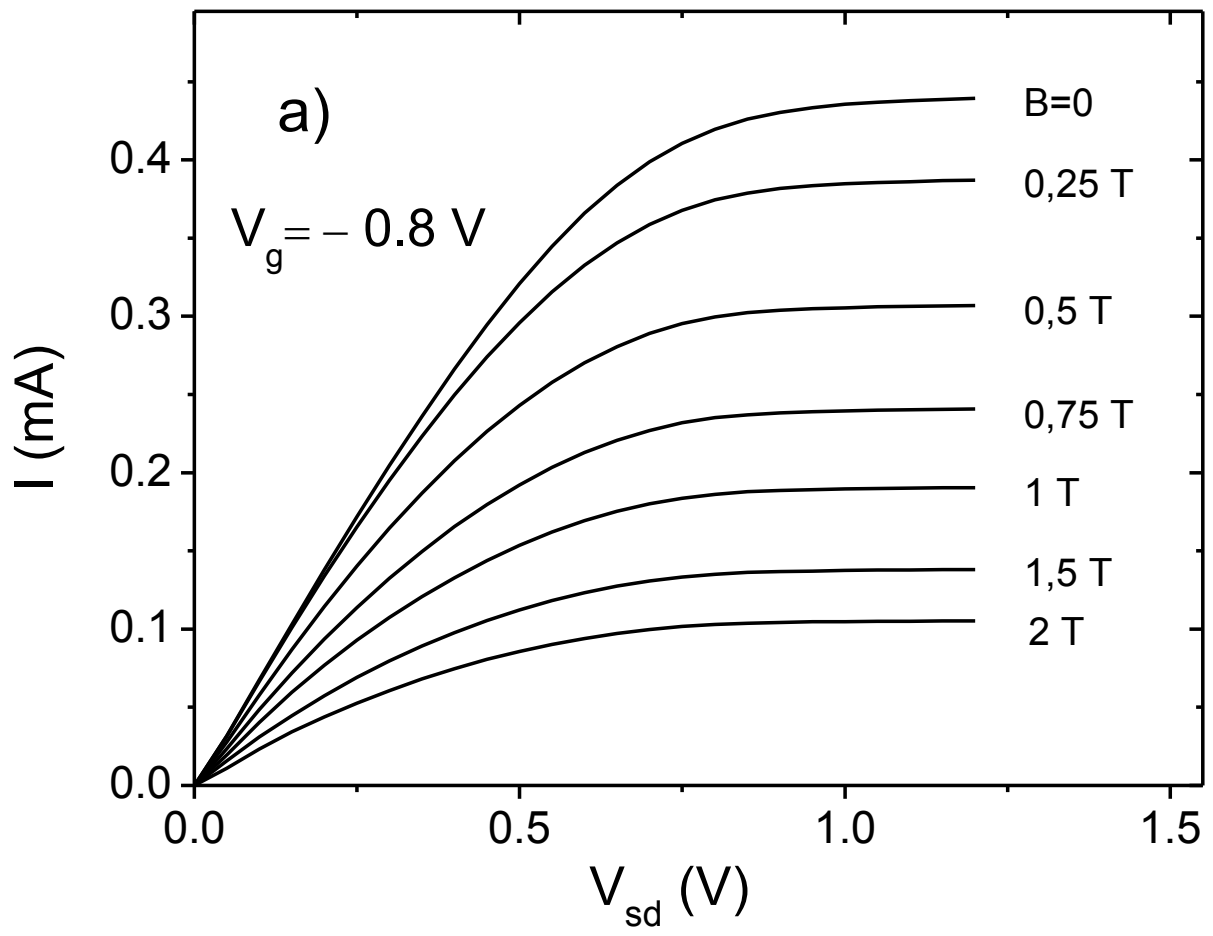

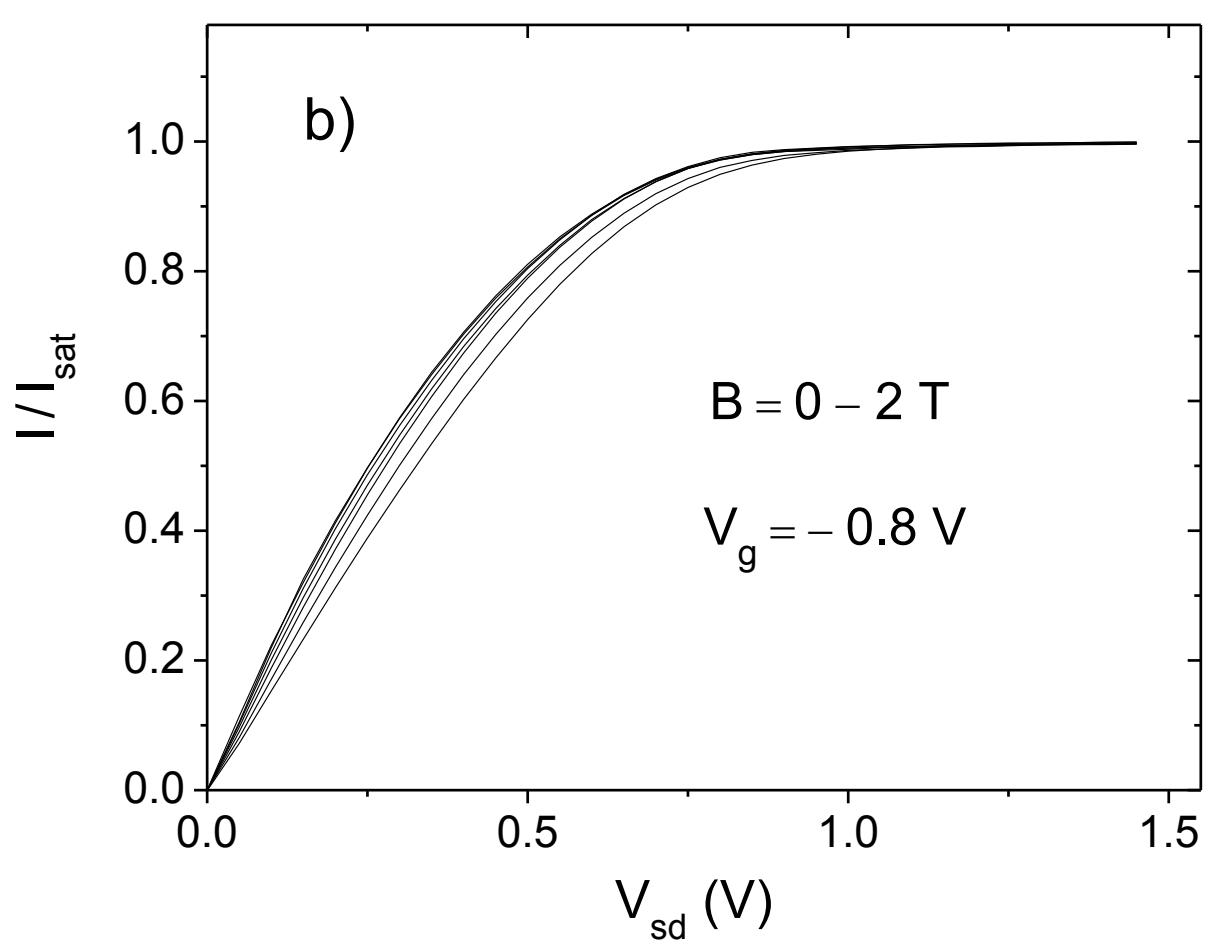

FIG. 6: (*a*) $I$ *vs* $V_{sd}$ dependencies for several values of magnetic field at $V_g$ = -0.8 *V*. (*b*) $I/I_{sat}$ *vs* $V_{sd}$ curves for several values of magnetic field and fixed $V_g$ = -0.8 *V*.

### *2. Hall effect*

In Fig. 7 we present the voltage differences $V_{31}$ and $V_{42}$ between edge contacts 3,1 and 4,2 respectively. In the linear regime when the electron concentration is homogeneous along the sample, these voltages are obviously equal because of the symmetry of the equipotential lines.

In the nonlinear regime, the equipotential lines correspond to fixed values of $V^2$ rather than of $V$, see Eq. (9). Thus, in the non-linear regime symmetry requires that $V_1^2 - V_3^2 = V_2^2 - V_4^2$. Hence

$$\frac{V_1 - V_3}{V_2 - V_4} = \frac{V_2 + V_4}{V_1 + V_3} \qquad (11)$$

This ratio is less than 1, because contacts 2 and 4 are closer to the drain than contacts 1 and 3 respectively (see Fig. 1) and thus correspond to regions where the electron concentration $n \sim V$ is lower. Also, as we have seen in the preceding section, the equipotential lines become frozen for $B$ > 1 T. Thus, in the nonlinear regime the equipotential lines of $V^2$ should remain unchanged above 1 T, resulting in saturation of all magnetic field effects.

These considerations explain qualitatively the results for Hall measurements in Fig. 7:

1) The difference between Hall voltages $V_{42}$ and $V_{31}$ appears in the nonlinear regime ($V_{42} > V_{31}$) and increases with $V_{sd}$ when approaching current saturation.

2) Both voltages $V_{42}$ and $V_{31}$ tend to saturate at $B$ > 1 T.

The measurements are in a good agreement with the predictions given by Eq. (10) (dashed lines in Fig. 7).

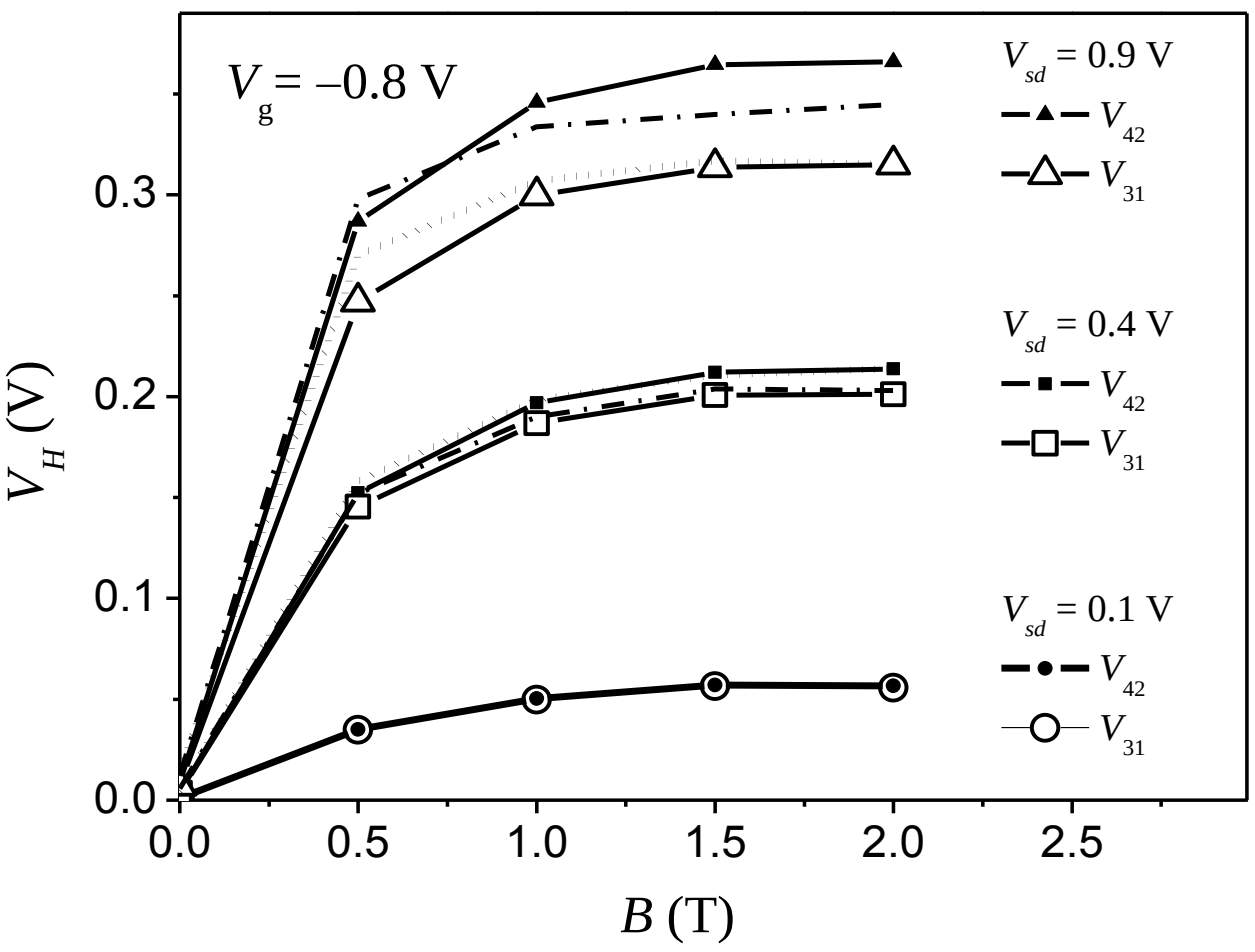

FIG. 7. The Hall voltage $V_H$ as a function of magnetic field measured between two pairs of contacts, 3, 1 and 4, 2 in the linear regime ($V_{sd}$ = 0.1V), weakly nonlinear regime ($V_{sd}$ = 0.4V) and strongly nonlinear regime ($V_{sd}$ = 0.9V). Dashed lines: calculation on the basis of results for the linear regime in Fig. 5, using Eq. (10).

The difference between the Hall voltages $V_{31}$ and $V_{42}$ in the nonlinear regime can be qualitatively understood as a result of the decrease of the electron concentration on the way from source to drain. Because of this, $V_{42} > V_{31}$.

### *3. Side voltages*

We have measured the voltage differences $V_{21}$ and $V_{43}$ between edge contacts 1, 2 and 3, 4 respectively. At zero magnetic field we have $V_{21} = V_{43}$ (see Fig. 1(*a*)). Because of the symmetry of equipotential lines, the same is true in the linear regime for any value of magnetic field. In the presence of magnetic field, the difference appears in the nonlinear regime. Similar to Eq. (11), we now have

$$\frac{V_4 - V_3}{V_2 - V_1} = \frac{V_2 + V_1}{V_4 + V_3} \qquad (12)$$

Depending on the sign of magnetic field, this ratio is either greater or smaller than 1. In our experimental configuration, the upper edge of the sample has a potential closer to that of the source $V_s$, while the lower edge has the potential closer to that of the drain $V_d$ (see Fig. 2(*b*)). Thus, in the nonlinear regime: $V_{43}/V_{21}>1$. We note that, in contrast to the relation between Hall voltages considered above, this qualitative result *a priori* is not evident.

Fig. 8 shows the side voltages $V_{21}$ and $V_{43}$ at $B$=0 and at $B = 2\ T$ as functions of $V_{sd}$. As we approach the nonlinear regime, those voltages start to diverge with increasing magnetic field. The inset shows the measured ratio $V_{43}/V_{21}$ as a function of magnetic field for several values of $V_{sd}$ together with the results of calculations using Eq. (10) and our experimental data for the linear regime in Fig. 5. One can again see a good agreement between experiment and theory. This shows that the difference between the real equipotential pattern accounting for the ungated access zones and the idealized pattern in Fig. 1 has no significant consequences.

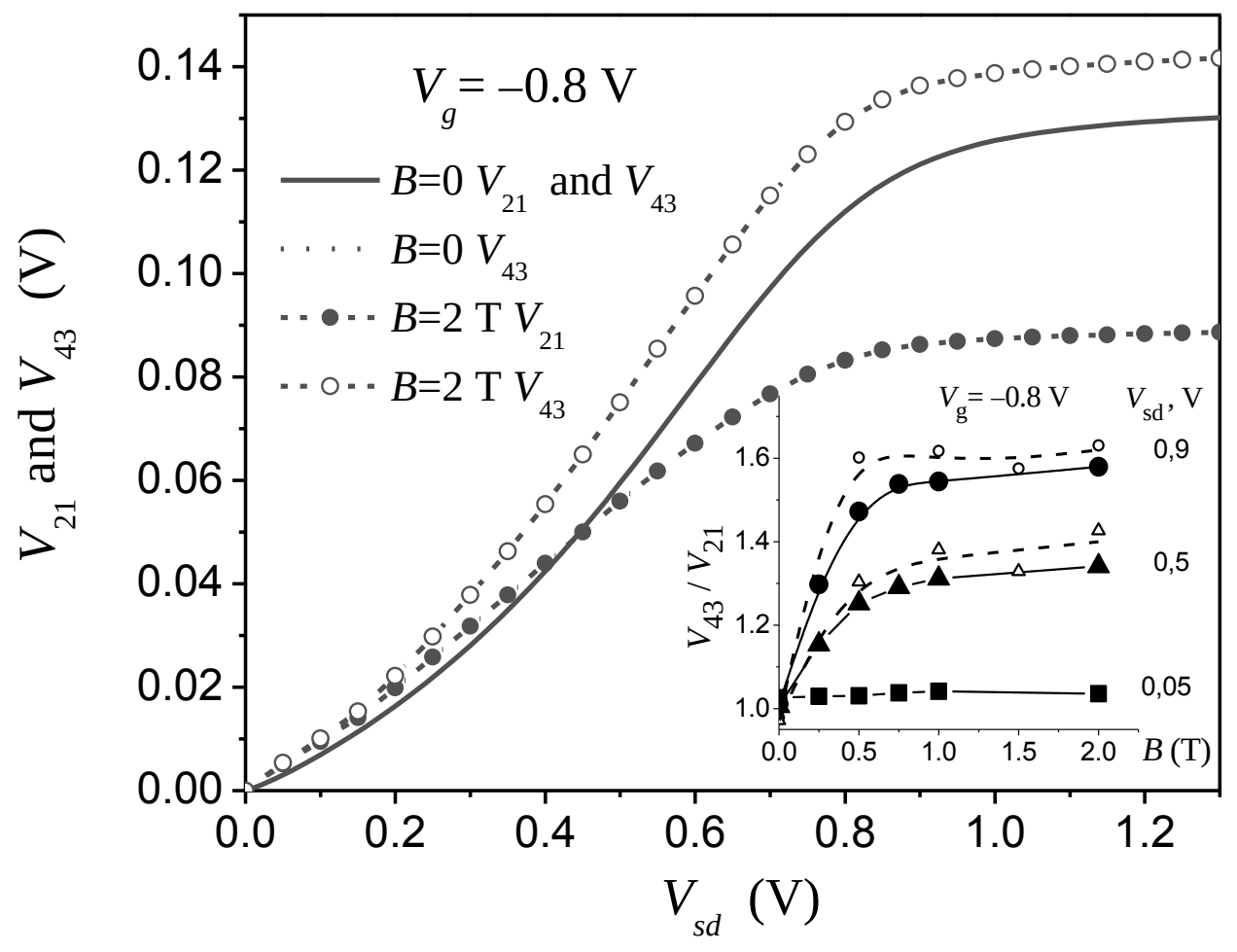

FIG. 8. Measured side voltages $V_{21}$ and $V_{43}$ as functions of the source-drain voltage $V_{sd}$. Solid line - at $B$=0. Open and full circles - at $B$=2 T. Inset: Full symbols and solid lines: measured ratio $V_{43}$ /$V_{21}$ as a function of magnetic field for several values of $V_{sd}$. Open symbols and dashed lines: calculation on the basis of results for the linear regime in Fig. 5, using Eq. (10).

## IV. CONCLUSIONS

In summary, we have measured the magnetoresistance, the Hall effect, and the side voltages in a gated 2D electron gas in magnetic field. The experimental results obtained in the nonlinear regime, when approaching the drain current saturation, have been interpreted based on the results obtained in the linear regime by a simple theoretical model, which describes quite well our observations.

## ACKNOWLEGMENTS

We thank Vladimir Umansky for very helpful discussions. At Laboratory Charles Coulomb this work was supported by CNRS through IRP TeraMIR project.

### References

[1] D.C. Look, J. Electrochem. Soc., **137,** 260 (1990).
[2] J.R. Drabble, R. Wolfe, J. Electronics and Control, **3**, 259 (1957).
[3] H.H. Jensen and H. Smith, J. Phys. C, **5**, 2867 (1972).
[4] N. Dyakonova, F. Teppe, J. Lusakowski, W. Knap, M. Levinshtein, A.P. Dmitriev, M.S. Shur, S. Bollaert and A. Cappy, J. of Appl. Phys. **97**, 114313 (2005).
[5] A.M. Chang and D.C. Tsui, Solid State Comm. **56**, 153 (1985).
[6] H.L. Stormer, K.W. Baldwin, L.N. Pfeiffer, and K.W. West, Solid State Comm. **84**, 95 (1992).
[7] V. Renard, Z.D. Kvon, G.M. Gusev, and J.C. Portal, Phys. Rev. B **70**, 033303 (2004).
[8] V.A. Johnson and W.J. Whitesell, Phys. Rev. **89**, 941 (1953).
[9] E.M Baskin, L.N. Magarill, and M.V. Entin, JETP **48**, 365 (1978).
[10] D.G. Polyakov, F. Evers, A.D. Mirlin, and P . Wölfle, Phys. Rev. B **64**, 205306 (2001).
[11] J. Ping, I. Yudhistira, N. Ramakrishnan, S.Cho, S. Adam, and M. Fuhrer, Phys. Rev. Lett. **113**, 047206 (2014).
[12] S. M. Sze and Kwok K. Ng, *Physics of Semiconductor Devices* (John Wiley and Sons, Inc, New Jersey, 2006).